\documentclass[epsfig,12pt,onecolumn]{article}
\usepackage{amsfonts}
\usepackage{amssymb}
\usepackage{amsmath}
\usepackage{multicol}
\usepackage{graphicx}
\usepackage{float}
\usepackage{caption}

\makeatletter \@addtoreset{equation}{section}

\def \be{\begin{equation}}
\def \ee{\end{equation}}
\def \bea{\begin{eqnarray}}
\def \eea{\end{eqnarray}}

\newcommand{\nc}{\newcommand}
\nc{\al}{\alpha} \nc{\bib}{\bibitem} \nc{\la}{\lambda}
\nc{\C}{\mbox{\hspace{1.24mm}\rule{0.2mm}{2.5mm}\hspace{-2.7mm} C}}
\nc{\R}{\mbox{\hspace{.04mm}\rule{0.2mm}{2.8mm}\hspace{-1.5mm} R}}
\input{tcilatex}
\begin{document}

\title{\textbf{Klein-Gordon and Schr\"{o}dinger solutions in Lovelock
quantum gravity}}
\author{M. Bousder$^{1}$\thanks{%
mostafa.bousder@fsr.um5.ac.ma}, A.Riadsolh$^{3}$, A. El Fatimy$^{4,5}$, \and %
M. El Belkacemi$^{3}$ and H. Ez-Zahraouy$^{1,2}$ \\
$^{1}${\small Laboratory of Condensend Matter and Interdisplinary Sciences,
Department of physics,}\ \\
{\small Faculty of Sciences, Mohammed V University in Rabat, Morocco}\\
$^{2}${\small CNRST Labeled Research Unit (URL-CNRST), Morocco}\\
$^{3}${\small Laboratory of Conception and Systems (Electronics, Signals and
Informatics)}\\
\ {\small Faculty of Sciences, Mohammed V University in Rabat, Morocco}\\
$^{4}${\small Central European Institute od Technology,}\\
\ {\small CEITEC BUT, Purky\v{n}ova 656/123, 61200 Brno, Czech Republic}\\
$^{5}${\small Departement of Physics, Universit\'{e} Mohammed VI
Polytechnique, Ben Guerir 43150, Morocco}}
\maketitle

\begin{abstract}
This study investigates the application of wave functions to explore various
solutions of the Klein-Gordon and Schr\"{o}dinger equations within the
framework of Lovelock gravity. We also present the derived Smarr formula
from the topological density. The Klein-Gordon solution leads to the
Wheeler-de Witt Hamiltonian and quasinormal modes, and we demonstrate the
connection between the potential and the black hole temperature within the
Schwarzschild limit. Additionally, we discuss different solutions of the Schr%
\"{o}dinger equation, with one solution highlighting the influence of the
Airy solution on the wave function's evolution over time.
\end{abstract}

\section{Introduction}

Lovelock gravity \cite{Z13} is an extension of Einstein's general theory of
relativity to higher dimensions \cite{JHEP1,CQG}. Recent advancements in
Lovelock theories have been proposed in \cite{PRD1,ANL1,ANLA1,PRD2,ANL2},
particularly the involvement of Lovelock gravity in the context of black
holes \cite{LOS1,LOS2,LOS3,LOS4,LOS5,LOS6,LOS7,LOS8}. Instead of insisting
on complete cancellation of the boundary term, Gibbons and Hawking
(motivated by quantum cosmology) proposed adding a specific counter-term
(now called the Gibbons-Hawking term) to exactly cancel the unwanted
variations. Gibbons and Hawking challenged the strictness of the earlier
requirement, suggesting a precise cancellation via an additional boundary
term (the Gibbons-Hawking term) in line with quantum cosmology.\cite%
{LOS9,LOS10}. Beyond its role in gravity, pseudo-entropy shines as a quantum
information tool, quantifying entanglement in states bridging initial and
final points \cite{AJ8,AJ9,LOS11}. Pseudo-entropy's significance lies in its
ability to quantify the entanglement present in intermediate states,
providing a quantitative metric for the intricate correlations that arise
during the process of evolution from one state to another. In the context of
quantum gravity, the customary approach treats the entire universe as a
quantum system \cite{QU0,QU1} often involving the contemplation of multiple
co-existing and non-interacting universes \cite{AnnQU,AnnQU2}. the authors
in \cite{PRLL1} have used the complexity=action conjecture to investigate
the late time growth of black hole complexity within Lovelock gravity.
Within a quantum mechanical framework, the measurement of complexity can be
achieved through Krylov Complexity \cite{PRX00,LOS12}, often referred to as
K-complexity. The computation is made feasible by the negligible
contribution of null boundaries from the Wheeler-DeWitt patch \cite{HC2,HC3}
at late-times, while essential contributions from the joints are now
well-understood. We recall that the Wheeler-DeWitt equation \cite{WD}, $%
H\Psi =0$, governing the wave function of the universe $\Psi $, is expressed
as a Schr\"{o}dinger equation within the Friedman-Robertson-Walker (FRW)
spacetime. An interesting inflationary model based on Lovelock terms is
discussed in the article \cite{LF1}. These terms, incorporating higher-order
curvature, can lead to inflation when there are more than three spatial
dimensions. The holographic principle states that all physics within
spacetime is encoded at its boundaries \cite{HC6}. Within the framework of
AdS/CFT, considerable knowledge exists regarding this encoding, particularly
for the region outside horizons. However, our understanding of the
holographic representation of black hole horizons remains limited. The
proposed alteration ensures that the universal component of holographic
subregion complexity maintains a proportional relationship with that of
holographic entanglement entropy \cite{WB4,WB3,WB0}, mirroring the behavior
in supergravity solutions lacking warp factors \cite{WB2}. This alignment
signifies a pivotal discovery, establishing a uniform leading behavior at
large N for both these quantities. Furthermore, this correlation between
universal components unveils intriguing "universal" relationships. These
connections extend to the field theoretical counterpart of holographic
subregion complexity, suggesting compelling links with the sphere partition
function or Weyl a-anomaly in odd or even dimensions, respectively. Through
rigorous analytical methods \cite{WB1}, an exact solution to the
Klein-Gordon equation is obtained, revealing quasinormal modes of a
distinctly imaginary nature. This peculiar characteristic signifies a lack
of oscillatory behavior, ensuring the stability of these perturbations. The
investigation extends further to examine crucial parameters such as greybody
factors, absorption cross-section, and decay rate for the non-Abelian
charged Lifshitz black branes. Through precise analytical derivations, the
researchers shed light on these fundamental aspects, facilitating a deeper
understanding of the strongly coupled dual theory. In the work \cite{WB} we
studied solutions for inflation and late-time cosmic acceleration within the
framework of quantum Lovelock gravity, employing Friedmann equations as the
basis. The authors explores hypergeometric states of cosmic acceleration by
delving into the Schr\"{o}dinger stationary equation. Additionally, the
paper offers predictions for the spectral tilt and tensor-to-scalar ratio
through plotted curves. By employing the rescaled Hubble parameter, the
spectral index is determined in relation to the number of e-folds. These
findings mark a significant advancement in our understanding of cosmic
acceleration within the quantum Lovelock gravity framework.\newline
The aim of this study is to find an interpretation of Lovelock coupling
following a connection between Lovelock gravity and quantum mechanics by
examining the solutions of the Klein-Gordon equation in terms of topological
density.\newline
The structure of this paper is outlined as follows: Section 2 focuses on
deriving topological densities from the equations of motion for Lovelock
gravity. Section 3 is dedicated to the study of solutions to the
Klein-Gordon equation. Section 4 is dedicated to the study of the solution
of the Schr\"{o}dinger equation. Finally, section 5 concludes the paper by
summarizing our findings.\textbf{\ }Throughout this article, we use units $%
c=\hbar =k_{B}=1$.

\section{Topological density}

\subsection{General Lovelock gravity}

Let us start with the Lagrangian of the $(D+1)$-dimensional Lovelock gravity
\cite{Z13} of extended Euler densities is given by%
\begin{equation}
\mathcal{L}=\sum_{k=0}^{\bar{D}}\frac{\alpha _{k}}{16\pi G_{N}}2^{-k}\delta
_{c_{1}d_{1}\cdots c_{k}d_{k}}^{a_{1}b_{1}\cdots a_{k}b_{k}}R_{a_{1}b_{1}}^{%
\text{ \ \ \ \ }c_{1}d_{1}}\ldots R_{a_{k}b_{k}}^{\text{ \ \ \ \ }%
c_{k}d_{k}},  \label{a1}
\end{equation}%
where $G_{N}$ is Newton's constant in $\left( D+1\right) $-dimensional
space-time. Here, $\bar{D}=\left[ \frac{D-1}{2}\right] $ denote taking the
integer part of the respective dimension $D$, $\delta _{c_{1}d_{1}\cdots
c_{k}d_{k}}^{a_{1}b_{1}\cdots a_{k}b_{k}}$\ is the generalized antisymmetric
Kronecker delta, $\alpha _{k}$ are the Lovelock coupling constants with
dimensions $(length)^{2k-D}$, and $R_{a_{1}b_{1}}^{\text{ \ \ \ \ }%
c_{1}d_{1}}$ is the generalized Riemann tensor. Here, $\alpha _{2}\ $and $%
\alpha _{3}$ are respectively the second (Gauss-Bonnet coupling \cite%
{JCAP4,PLB2}) and the third Lovelock coefficients. We introduce the
effective thermodynamic pressure \cite{CQG2} as
\begin{equation}
p=\frac{\alpha _{0}}{16\pi G_{N}}=-\frac{\Lambda }{8\pi G_{N}}=\frac{\left(
D-1\right) \left( D-2\right) }{16\pi G_{N}L^{2}},  \label{ap}
\end{equation}%
with $\Lambda $ being the cosmological constant and $L$ is the AdS radius
\cite{JHEAP1}. The equations of motion for Lovelock gravity \cite{CQG4} read%
\begin{equation}
\sum_{k=0}^{\bar{D}}\frac{\alpha _{k}}{16\pi G_{N}}2^{-2k-1}\delta _{\mathbf{%
a}a_{1}b_{1}\cdots a_{k}b_{k}}^{\mathbf{c}c_{1}d_{1}\cdots
c_{k}d_{k}}R_{c_{1}d_{1}}^{\text{ \ \ \ \ }a_{1}b_{1}}\ldots =0.
\end{equation}%
Let us recall the expression of the ansatz%
\begin{equation}
ds^{2}=-f(r)dt^{2}+\frac{1}{f(r)}dr^{2}+r^{2}d\Omega _{D-2}^{2}.  \label{ds1}
\end{equation}%
The corresponding solution is given by the metric function \cite{CQG}:
\begin{equation}
f(r)=\kappa +r^{2}\left( \bar{D}\alpha \right) ^{-\frac{1}{K-1}}\left(
1-\left( 1+\frac{m\left( r\right) -\alpha _{0}}{\alpha \left( \bar{D}\alpha
\right) ^{-\frac{K}{K-1}}}\right) ^{\frac{1}{\bar{D}}}\right) ,  \label{a7}
\end{equation}%
with $\kappa $ is the horizon curvatures with $\kappa =\left\{ 0,\pm
1\right\} $ for\ $\Lambda <0$ and $\kappa =+1$ for\ $\Lambda \geq 0$. Thus
we can uniformly rewrite the Lovelock equations as the Arnowitt-Deser-Misner
(ADM) mass $M$ of black hole \cite{NPB1}:%
\begin{eqnarray}
m(r) &=&\frac{16\pi G_{N}M}{\left( D-2\right) \Sigma _{D-2}r^{D-1}}
\label{a5} \\
&=&\frac{\alpha _{0}}{\left( D-1\right) \left( D-2\right) }+\sum_{k=0}^{\bar{%
D}}\alpha _{k}\left( \frac{\kappa -f(r)}{r^{2}}\right)
^{k}\prod\limits_{l=3}^{2k}\left( D-l\right) ,
\end{eqnarray}%
where $\Sigma _{D-2}$ is the area of a radius $(D-2)$-dimensional unit
sphere: $\Sigma _{D-2}=\frac{2\pi ^{\frac{\left( D-1\right) }{2}}}{\Gamma
\left( \frac{D-1}{2}\right) }.$ A standard approach involves examining the
relationship between the black hole area which is an intrinsic property of
the horizon and dynamical quantities (total energy or angular momentum...).
In an asymptotically AdS black hole spacetime, the black hole thermodynamic
volume with $\left( D-1\right) $-dimensions, derived entirely from
thermodynamic considerations. This volume is thermodynamically conjugate to
the pressure $P$ \cite{CQG7}. The $\left( D-1\right) $-dimensional volume of
the black hole of radius $r_{H}$ is%
\begin{equation}
V_{D-1}=\frac{2\pi ^{\frac{\left( D-1\right) }{2}}}{\left( D-1\right) \Gamma
\left( \frac{D-1}{2}\right) }r_{H}^{D-1}=\frac{\Sigma _{D-2}r_{H}^{D-1}}{%
\left( D-1\right) },  \label{a6}
\end{equation}%
Here, $r_{H}$\ denotes the radius of the black hole is one of the roots of
the metric function Eq. (\ref{a7}). The black hole entropy \cite{CQG} is
given by%
\begin{equation}
S=\frac{\Sigma _{D-2}\left( D-2\right) \alpha }{4G_{N}}\sum_{k=1}^{\bar{D}%
}\left(
\begin{array}{c}
\bar{D} \\
k%
\end{array}%
\right) \frac{k\kappa ^{k-1}r_{H}^{D-2k}}{\left( D-2k\right) \left( \bar{D}%
\alpha \right) ^{\frac{\bar{D}-k}{\bar{D}-1}}}.  \label{sl}
\end{equation}%
Using the D-dimensional static solutions in the Lovelock gravity for the
rescaled Lovelock coupling $\alpha $, we have%
\begin{equation}
\alpha _{k}=\alpha \left( \bar{D}\alpha \right) ^{-\frac{\bar{D}-k}{\bar{D}-1%
}}\left(
\begin{array}{c}
\bar{D} \\
k%
\end{array}%
\right) ,2\leq k\leq \bar{D}.
\end{equation}%
The event horizon in spacetime can be located by solving the metric
equation: $f(r)=0$, which yields%
\begin{equation}
\frac{m\left( r_{H}\right) }{\alpha \left( \bar{D}\alpha \right) ^{-\frac{%
\bar{D}}{\bar{D}-1}}}=-1+\frac{\alpha _{0}}{\alpha \left( \bar{D}\alpha
\right) ^{-\frac{\bar{D}}{\bar{D}-1}}}+\left[ 1+\kappa \frac{\left( \bar{D}%
\alpha \right) ^{\frac{1}{\bar{D}-1}}}{r_{H}^{2}}\right] ^{\bar{D}}.
\label{a8}
\end{equation}%
One should note that in the event horizon $r_{H}$ the mass function is given
by $m(r_{H})=\frac{16\pi G_{N}M}{\left( D-2\right) \Sigma _{D-2}r_{H}^{D-1}}$%
, so that one obtains the following solution%
\begin{equation}
\frac{16\pi G_{N}\left( \bar{D}\alpha \right) ^{\frac{\bar{D}}{\bar{D}-1}%
}\rho _{BH}}{\alpha \left( D-2\right) \left( D-1\right) }=-1+\frac{\alpha
_{0}}{\alpha }\left( \bar{D}\alpha \right) ^{\frac{\bar{D}}{\bar{D}-1}}+%
\left[ 1+\kappa \frac{\left( \bar{D}\alpha \right) ^{\frac{1}{\bar{D}-1}}}{%
r_{H}^{2}}\right] ^{\bar{D}}.  \label{a10}
\end{equation}%
Here $\rho _{BH}=M/V_{D-1}$ is the $\left( D-1\right) $-dimensional black
hole density. Using the binomial theorem
\begin{equation}
\left[ 1+\frac{\kappa \left( \bar{D}\alpha \right) ^{\frac{1}{\bar{D}-1}}}{%
r^{2}}\right] ^{\bar{D}}=\sum_{k=0}^{\bar{D}}\left(
\begin{array}{c}
\bar{D} \\
k%
\end{array}%
\right) \left( \frac{\kappa \left( \bar{D}\alpha \right) ^{\frac{1}{\bar{D}-1%
}}}{r^{2}}\right) ^{k}.  \label{a11}
\end{equation}%
By considering a specific pressure of solutions corresponding to the choice
of the equation of state (EoS) parameter of dark energy $\omega _{\Lambda
}=- $ $p/\rho _{\Lambda }$. From Eqs. (\ref{a10})-(\ref{a11}) the general
solution can be written as%
\begin{equation}
\frac{\rho _{BH}}{\left( D-2\right) \left( D-1\right) }=p+\frac{\kappa }{3}%
\rho _{\Lambda }+\alpha \left( \bar{D}\alpha \right) ^{-\frac{\bar{D}}{\bar{D%
}-1}}\sum_{k=2}^{\bar{D}}\left(
\begin{array}{c}
\bar{D} \\
k%
\end{array}%
\right) \frac{\left( \kappa \left( \bar{D}\alpha \right) ^{\frac{1}{\bar{D}-1%
}}\right) ^{k}}{16\pi G_{N}r_{H}^{2k}},  \label{a12}
\end{equation}%
where $\rho _{\Lambda }\equiv \frac{3}{16\pi G_{N}r_{H}^{2}}$ is the density
of the holographic dark energy \cite{D9,ANL1}. In this case we introduce the
topological density $\rho _{\alpha }$ as%
\begin{equation}
\rho _{\alpha }=\frac{\alpha }{16\pi G_{N}}\sum_{k=2}^{\bar{D}}\left(
\begin{array}{c}
\bar{D} \\
k%
\end{array}%
\right) \kappa ^{k}\frac{\left( \bar{D}\alpha \right) ^{\frac{k-\bar{D}}{%
\bar{D}-1}}}{r_{H}^{2k}}.  \label{a13}
\end{equation}%
This topological density depends directly on the Lovelock coupling.

\subsection{4d topological density}

We set $\kappa =\pm 1$ and $D=4$ and we get the 4-dimensional topological
density%
\begin{equation}
\rho _{\alpha }=\frac{\alpha }{16\pi G_{N}}\frac{1}{r_{H}^{4}}.  \label{a14}
\end{equation}%
Now consider the standard Hawking temperature $T=\frac{1}{8\pi G_{N}r_{H}}$,
therefore $\rho _{\alpha }=3\alpha \left( 8\pi G_{N}\right) ^{3}T^{4}$. From
the Stefan--Boltzmann law, the topological density $\rho _{\alpha }\propto
T^{4}$ represents system radiation. From (\ref{a12})-(\ref{a14}) we have $%
\rho _{BH}=6p+\kappa \frac{3}{8\pi G_{N}r_{H}^{2}}+\frac{3\alpha }{8\pi
G_{N}r_{H}^{4}}$, or equivalently%
\begin{equation}
\rho _{BH}=6p+3\kappa \left( 8\pi G_{N}\right) T^{2}+3\alpha \left( 8\pi
G_{N}\right) ^{3}T^{4}.  \label{a17}
\end{equation}%
We can conclude from (\ref{a17}) the Van der Waals equation $\left( \rho
_{BH}-6p\right) V=\left( 4\pi r_{H}^{2}\kappa +4\pi \alpha \right) T$. The
equation of state $\omega _{H}=p/\rho _{BH}$ at the horizon can be
summarized as:%
\begin{equation}
\left( 1-6\omega _{H}\right) \frac{M}{T}=4\pi r_{H}^{2}\kappa +4\pi \alpha .
\label{a18}
\end{equation}%
We notice that if $\omega _{H}=1/6$, we obtain $\alpha =-\kappa r_{H}^{2}$.
This finding demonstrates a consistent correlation between the topological
termination and the spatial parameter "$r$". Furthermore, as stated in (\ref%
{a18}), we note that the term $\frac{M}{T}$ corresponds to the entropy, in
accordance with the black hole first law. Within the limit $\alpha
\rightarrow 0$, we find the Bekenstein-Hawking entropy $S_{BH}=\frac{\pi
r_{H}^{2}}{G_{N}}$. We note that Bekenstein-Hawking entropy is equivalent to
$\omega _{H}=-\frac{1}{3}$ or $\omega _{H}=\frac{5}{6}$. The limit $\alpha
\rightarrow 0$ is equivalent to $\omega _{H}=\frac{1-4\kappa }{6}$. The
entropy $S_{BH}$ in (\ref{a18}) is connected with the Smarr formula \cite%
{CQG7}:%
\begin{equation}
\left( \frac{1-6\omega _{H}}{4\kappa }\right) M=G_{N}TS_{BH}+\frac{\pi
\alpha }{\kappa }T.  \label{a20}
\end{equation}%
In addition, two temperatures are obtained $\beta =\frac{4\pi }{M}\frac{%
\alpha \pm r_{H}^{2}}{1-6\omega _{H}}$. The temperature is influenced by the
topological structure indicated by Lovelock coupling, as we have observed.
From (\ref{a14}) so that one obtains the following solution $M\approx \frac{%
\alpha }{12G_{N}r_{S}}$. In this particular case, the Schwarzschild radius $%
r_{S}$ is solely determined by the Lovelock coupling as $r_{S}^{2}\approx
\frac{\alpha }{6}$.

\section{Lovelock-Klein-Gordon solutions}

\subsection{Hamiltonian of Wheeler-DeWitt equation}

Lovelock gravity is associated with $(D+1)$ dimensional AdS, while quantum
states reside within $D$ dimensions. First we will study the solutions in
space-time assuming that $f^{2}(r)\sim 1$. Secondly we will compare this
solution with that of space time curved according to the tortoise
coordinates. In this scenario, we can formulate the Klein Gordon equation
for $D$ dimensions and quantum states that exist in $(D+1)$ dimensions.%
\begin{equation}
\left( g^{\mu \nu }\nabla _{\mu }\nabla _{\nu }+M^{2}\right) \Psi =0,
\label{aa1}
\end{equation}%
where $\Psi $ is the wave function of the universe. In the expression of Eq.
(\ref{a13}) we have the\ topological density of the black hole, and
according to Eq. (\ref{a12}) the density of the black hole is expressed in
the form of the sum $\sim p+\frac{\kappa }{3}\rho _{\Lambda }+\rho _{\alpha
} $, and if we neglect the pressure $p$\ and the density $\rho _{\Lambda }$\
in front of $\rho _{\alpha }$, so we suggest that $M\equiv \left( p+\frac{%
\kappa }{3}\rho _{\Lambda }+\rho _{\alpha }\right) V_{D-1}\approx \rho
_{\alpha }V_{D-1}$. Consequently from Eqs. (\ref{a13},\ref{aa1}), we derive
the Klein-Gordon equation in Lovelock gravity for choice $g^{\mu \nu }\nabla
_{\mu }\nabla _{\nu }=\frac{\partial ^{2}}{\partial t^{2}}-\frac{\partial
^{2}}{\partial r^{2}}$:%
\begin{equation}
\left( \frac{\partial ^{2}}{\partial t^{2}}-\frac{\partial ^{2}}{\partial
r^{2}}+\frac{\alpha ^{2}V_{D-1}^{2}}{\left( 16\pi G_{N}\right) ^{2}}\left(
\sum_{k=2}^{\bar{D}}\left(
\begin{array}{c}
\bar{D} \\
k%
\end{array}%
\right) \kappa ^{k}\frac{\left( \bar{D}\alpha \right) ^{\frac{k-\bar{D}}{%
\bar{D}-1}}}{r_{H}^{2k}}\right) ^{2}\right) \Psi =0.  \label{aa2}
\end{equation}%
We know that the Hamiltonian of Wheeler-DeWitt equation \cite{AJ10,AJ11}
represents the quantum gravity verified $H_{D}\Psi =0$. The Hamiltonian $%
H_{D}$, represents the quantized version of the general relativity
Hamiltonian \cite{WDJ1,WDJ2,WDJ3,WDJ4,WDJ5}. So according to (\ref{aa2}), we
have%
\begin{equation}
\mathcal{H}_{D}=\frac{\partial ^{2}}{\partial t^{2}}-\frac{\partial ^{2}}{%
\partial r^{2}}+\frac{\alpha ^{2}V_{D-1}^{2}}{\left( 16\pi G_{N}\right) ^{2}}%
\left( \sum_{k=2}^{\bar{D}}\left(
\begin{array}{c}
\bar{D} \\
k%
\end{array}%
\right) \kappa ^{k}\frac{\left( \bar{D}\alpha \right) ^{\frac{k-\bar{D}}{%
\bar{D}-1}}}{r_{H}^{2k}}\right) ^{2}.  \label{aa3}
\end{equation}%
If $\alpha =0$\ we obtain $H_{0}=\frac{\partial ^{2}}{\partial t^{2}}-\frac{1%
}{f^{2}(r)}\frac{\partial ^{2}}{\partial r^{2}}$. The term multiply by $%
\alpha ^{2}$ refers to the topological curvature induced by Lovelock
gravity, which shows this Hamiltonian describes the quantum aspect of
gravity\ \cite{QTG1,QTG2,QTG3}.

\subsection{$D\rightarrow 4$ solution}

\begin{figure}[H]
\centering\includegraphics[width=11cm]{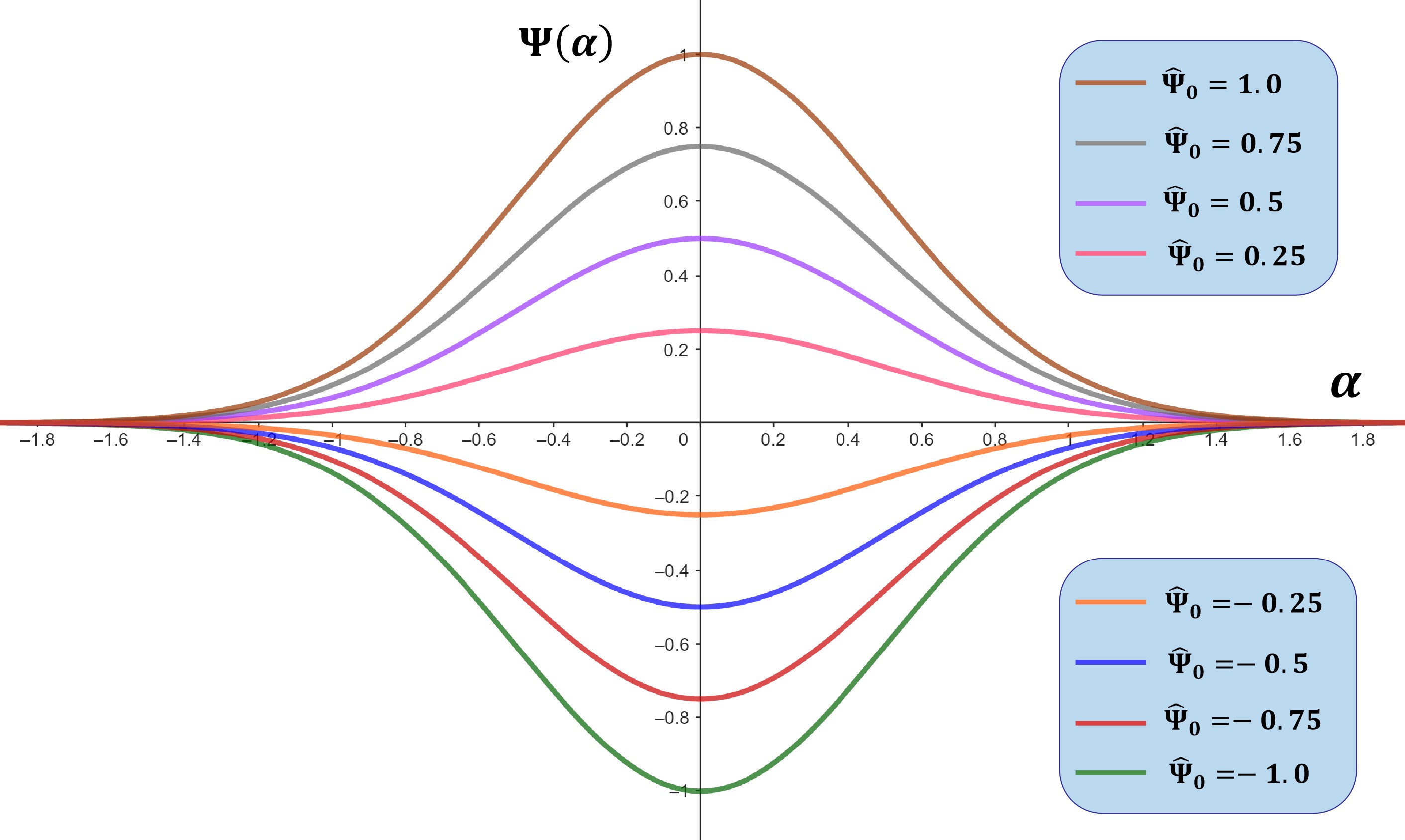}
\caption{The figure displays the wave function $\Psi \left( \protect\alpha %
\right) $ for different values of the scaling factor $\hat{\Psi}_{0}=\Psi
_{0}e^{\protect\lambda \left( t-r_{H}\right) }$ and a constant value of $r$
with $r=r_{H}=1/(G_{N}^{2}3^{2}2^{5})$. Let's choose specific values for
each of the parameters: $\Psi _{0}=1,$ $\protect\lambda =0.5,$$t=4,$ $r_{H}=2
$ and $r=3$. Using these values, we can calculate the function$\hat{\Psi}%
_{0}\approx 2.718$\textbf{.}}
\label{F1}
\end{figure}

Subsequently, we want to give a particular solution to the equation (\ref%
{aa2}). For $D\rightarrow 4$ and for $\nabla ^{2}=\partial _{r}^{2}=\partial
^{2}/\partial r^{2}$ and $\kappa =\pm 1$ we get%
\begin{equation}
\left( \frac{\partial ^{2}}{\partial t^{2}}-\frac{\partial ^{2}}{\partial
r^{2}}+\frac{\alpha ^{2}V_{4-1}^{2}}{\left( 16\pi G_{N}\right) ^{2}r_{H}^{8}}%
\right) \Psi \left( t,r\right) =0.  \label{ab1}
\end{equation}%
To solve this equation, we can use the method of separation of variables.
Let's assume that the solution can be expressed as a product of two
functions:%
\begin{equation}
\Psi \left( t,r\right) =\chi \left( t\right) \Phi \left( r\right) .
\label{ab2}
\end{equation}%
Now, let's substitute this into the Klein-Gordon equation, and we obtain two
ordinary differential equations to solve:%
\begin{eqnarray}
\dot{\chi}\left( t\right)  &=&\lambda \chi \left( t\right) ,  \label{ab3} \\
\frac{\partial \Phi \left( r\right) }{\partial r} &=&-\left( \lambda +\frac{%
\alpha ^{2}V_{4-1}^{2}}{\left( 16\pi G_{N}\right) ^{2}r_{H}^{8}}\right) \Phi
\left( r\right) ,  \notag
\end{eqnarray}%
where $\lambda $ is a constant. The volume of $D\rightarrow 4$ is $%
V_{4-1}=4\pi r^{3}/3$. Therefore, the general solution to the Klein-Gordon
equation is $\Psi \left( \alpha ,r,t\right) =\Psi _{0}e^{\lambda \left(
t-r\right) }\exp (-\frac{\alpha ^{2}r^{7}}{3^{2}2^{4}G_{N}^{2}r_{H}^{8}})$,
where $\Psi _{0}$ is an arbitrary constant Fig. (\ref{F1}).
Near the horizon we have $\log \left( \frac{\Psi _{H}\left( \alpha ,t\right)
}{\Psi _{0}}\right) =\lambda \left( t-r_{H}\right) -\frac{\alpha ^{2}}{%
12^{2}G_{N}^{2}r_{H}}$ which leads to%
\begin{equation}
\frac{\pi r_{H}}{\lambda G_{N}}\log \left( \frac{\Psi _{H}\left( \alpha
,t\right) }{\Psi _{0}}\right) =\frac{\pi r_{H}}{G_{N}}t-\frac{\pi r_{H}^{2}}{%
G_{N}}-\frac{\pi \alpha ^{2}}{12^{2}\lambda G_{N}^{3}},  \label{ab6}
\end{equation}%
with $\Psi _{H}$ is the wave function near the horizon. The term $\frac{\pi
r_{H}^{2}}{G_{N}}$ is the Bekenstein-Hawking entropy $S_{BH}$. The Wick
rotation $t=-it_{E}$, leads to%
\begin{equation}
S_{BH}+\frac{\alpha ^{2}\pi }{12^{2}\lambda G_{N}^{3}}=-\frac{\pi r_{H}}{%
\lambda G_{N}}\log \left( \frac{\Psi _{H}\left( \alpha ,t\right) }{\Psi _{0}}%
\right) -\frac{i\pi r_{H}t_{E}}{G_{N}}.  \label{ab7}
\end{equation}%
Subsequently, we will study a particular case of the Eq. (\ref{a18}), this
case is determined by an equivalence between topology and space $r=\frac{%
\alpha }{\sqrt{\alpha _{0}}}$\ or $\partial _{r}^{2}\equiv \alpha
_{0}\partial _{\alpha }^{2}$\ near the horizon ($r\sim r_{H}$). In this
context, the wave function relies on two indispensable parameters $\left(
\alpha ,t\right) $. From Eq. (\ref{ab1}) in $D=4$\ and we get
\begin{equation}
\left( \partial _{t}^{2}-\alpha _{0}\partial _{\alpha }^{2}+\alpha ^{2}\frac{%
r^{6}}{9G_{N}^{2}16r_{H}^{8}}\right) \Psi \left( \alpha ,t\right) =0,
\label{ac2}
\end{equation}%
where $V(D=4)=4\pi r^{3}/3$. To solve this equation, we can again use the
method of separation of variables. Let's assume that the solution can be
expressed as a product of two functions: $\Psi \left( \alpha ,t\right) =\chi
\left( t\right) \Phi \left( \alpha \right) $. Now, let's substitute this
into Eq. (\ref{ac2}), and we obtain two ordinary differential equations to
solve:%
\begin{eqnarray}
\dot{\chi}\left( t\right)  &=&\lambda \chi \left( t\right) ,  \label{ac3} \\
\frac{\Phi ^{\prime }\left( \alpha \right) }{\Phi \left( \alpha \right) } &=&%
\frac{\alpha ^{2}V^{2}\kappa ^{4}}{\left( 16\pi G_{N}\right) ^{2}r_{H}^{8}}%
-\lambda ,  \label{ac4}
\end{eqnarray}%
where $\lambda $ is a constant and $\dot{\chi}=\partial _{t}\chi $. We can
solve these linear differential equations Eq. (\ref{ac3})-(\ref{ac4}) using
an integrating factor. Therefore, the solution is%
\begin{equation}
\Psi \left( \alpha ,t\right) =\Psi _{0}\exp \left( \lambda \left( t+\frac{%
\alpha }{\alpha _{0}}\right) -\frac{\alpha ^{3}r^{6}}{\alpha
_{0}G_{N}^{2}3^{3}2^{4}r_{H}^{8}}\right) .  \label{ac5}
\end{equation}%
Near the horizon, (\ref{ac5}) becomes%
\begin{equation}
\pi r_{H}^{2}\log \left( \frac{\Psi \left( \alpha ,t\right) }{\Psi _{0}}%
\right) =-\left( t+\frac{\alpha }{\alpha _{0}}\right) \frac{\lambda \alpha
^{3}}{3^{3}2^{4}\alpha _{0}G_{N}^{2}}.  \label{ac6}
\end{equation}%
The Wick rotation $t=-it_{E}$ and the Bekenstein-Hawking entropy $S_{BH}=%
\frac{\pi r_{H}^{2}}{G_{N}}$, leads to%
\begin{equation}
\frac{\alpha ^{2}}{\alpha _{0}^{2}}\frac{\lambda \alpha ^{2}}{%
3^{3}2^{4}G_{N}^{3}}=S_{BH}\log \left( \frac{\Psi _{0}}{\Psi \left( \alpha
,t\right) }\right) +\frac{i\pi }{6}\frac{\lambda \alpha ^{3}t_{E}}{72\alpha
_{0}\pi G_{N}^{3}}.  \label{ac8}
\end{equation}%
Using $\alpha _{0}=-2\Lambda $ for $\Psi =\Psi _{0}$ and we get $\alpha
\left( \Psi =\Psi _{0}\right) =2\Lambda t$. According to the Friedmann
equations, the term $\Lambda $ is proportional to the square of the Hubble
parameter $H^{2}$, i.e. $\alpha \left( \Psi =\Psi _{0}\right) \propto H^{2}$%
. We are now using tortoise coordinate for our analysis. The differential
form $dr$\ can be written in terms of the tortoise coordinate as $dr^{\ast }=%
\frac{dr}{f(r)}$. Hence, from Eq. (\ref{ds1})%
\begin{equation}
ds^{2}=f(r^{\ast })\left( dr^{\ast 2}-dt^{2}\right) +r^{2}d\Omega _{D-2}^{2}.
\label{ds2}
\end{equation}%
In this case the singularity is found in $r_{H}^{\ast }\rightarrow \infty $%
.\ So Eq. (\ref{aa2}) is written as%
\begin{equation}
\left( \frac{\partial ^{2}}{\partial t^{2}}-\frac{1}{f^{2}(r)}\frac{\partial
^{2}}{\partial r^{\ast 2}}\right) \Psi =0,  \label{ds3}
\end{equation}%
and the Hamiltonian is written in the following form%
\begin{equation}
\mathcal{H}_{D}=\frac{\partial ^{2}}{\partial t^{2}}-\frac{1}{f^{2}(r)}\frac{%
\partial ^{2}}{\partial r^{\ast 2}}.  \label{ds4}
\end{equation}%
The general solution of Eq. (\ref{ds3}) equations for the metric Eq. (\ref%
{ds2}), takes the following form%
\begin{equation}
\Psi \left( t,r^{\ast }\right) =\Psi _{-}\left( r^{\ast }-\frac{t}{f(r)}%
\right) +\Psi _{+}\left( r^{\ast }+\frac{t}{f(r)}\right) .  \label{ds5}
\end{equation}%
The parameter $r^{\ast }$\ absorbs the topological term $\alpha $, while the
other time term, $f(r)$, encodes both topological and geometric information.

\subsection{Spherical harmonics and temperature of the black hole}

The scalar field $\Psi $ describes the general state of gravity and
undergoing static and spherically symmetric geometry decomposition can be
written as%
\begin{equation}
\Psi \left( t,r\right) =\int_{-\infty }^{+\infty }\Psi \left( t,r,\omega
\right) e^{-i\omega t}d\omega .  \label{p1}
\end{equation}%
and $\Psi \left( t,r,\omega \right) $ is given by \cite{nPRD1},%
\begin{equation}
\Psi \left( t,r,\omega \right) =\frac{1}{r^{\left( D-2\right) /2}}%
\sum_{l=0}^{\infty }\sum_{m=-l}^{l}e^{-i\omega t}Y_{lm}\left( \omega \right)
\psi _{lm}\left( r\right) .  \label{p2}
\end{equation}%
Here, $\psi _{lm}$ is the radial function and $Y_{lm}\left( \omega \right) $
characterizes the spherical harmonics. The function $\psi _{lm}\left(
r\right) $ satisfies the following differential equation:%
\begin{equation}
\frac{d^{2}\psi _{lm}}{dr^{2}}+\left( \omega ^{2}-V_{l}\left( r\right)
\right) \psi _{lm}=0,  \label{p3}
\end{equation}%
and the potential $V_{l}\left( r\right) $ is given by%
\begin{equation}
V_{l}\left( r\right) =\frac{f(r)}{r}\left( \frac{l\left( l+D-3\right) }{r}+%
\frac{\left( D-2\right) \left( D-4\right) }{4r}f(r)+\frac{\left( D-2\right)
}{2}f^{\prime }(r)\right) .  \label{p4}
\end{equation}%
The temperature of the black hole corresponds to the $T=\frac{1}{4\pi }%
f^{\prime }(r_{H})$. So, we find
\begin{equation}
\frac{r}{f(r)}V_{l}\left( r\right) -\frac{\left( D-2\right) \left(
D-4\right) }{4r}f(r)=\frac{l\left( l+D-3\right) }{r}+2\pi \left( D-2\right)
T.  \label{p5}
\end{equation}%
This relationship linked well the temperature of the black hole with the
potential $V_{l}\left( r\right) $.One recovers the temperature of the
Schwarzschild black hole in the limit of $\left( V_{l}\left( r\right) =\frac{%
f(r)}{r^{2}},D=4\text{ and }l=0\right) $, in which cas $T=1/4\pi r_{H}$ as
expected. Using $dr^{\ast }=\frac{dr}{f(r)}$, the potential that corresponds
to the Schwarzschild limit can also be written in the following form $%
V_{l}\left( r\right) =\frac{d}{dr^{\ast }}\left( -\frac{1}{r}\right) $.

\section{Wave function from Airy function}

\subsection{Schr\"{o}dinger equation}

Regarding the equivalence of $\alpha $\ and $r$\ in the horizon in (\ref{ac2}%
): $r=\frac{\alpha }{\sqrt{\alpha _{0}}}$\ or $\partial _{r}^{2}\equiv
\alpha _{0}\partial _{\alpha }^{2}$, we introduce the wave function $\Psi
=\Psi \left( \alpha ,t\right) $ of quantum states around the horizon ($r\sim
r_{H}$) of the black hole. Consequently, we derive the Schr\"{o}dinger
equation (see Appendix A):%
\begin{equation}
i\frac{\partial \Psi }{\partial t}=-\frac{\alpha _{0}}{2M}\frac{\partial
^{2}\Psi }{\partial \alpha ^{2}}+V\Psi .  \label{b1}
\end{equation}%
It is equivalent to the proposal of $\Psi =\Psi \left( \alpha ,t\right) $
that assigns the cosmological scale factor $a(t)$ at each time $t$ in cosmic
acceleration \cite{in2}. Here, $\left\vert \Psi \right\vert ^{2}$ is the
formal probability amplitude to find a given black hole of mass $M$ at a
given Lovelock coupling $\alpha (t)$ at time $t$ \cite{in2}. We consider a
stationary state with energy $E$ as $\Psi \left( \alpha ,t\right) =\psi
\left( \alpha \right) e^{-iEt}$, and the Schr\"{o}dinger stationary equation
is
\begin{equation}
\frac{\partial ^{2}\psi \left( \alpha \right) }{\partial \alpha ^{2}}+2\frac{%
M}{\alpha _{0}}\left( E-V\right) \psi \left( \alpha \right) =0.  \label{b3}
\end{equation}%
There are two methods to establish a connection between this equation and
Lovelock gravity. The first method involves utilizing only the topological
density (\ref{a13}) to analyze the states associated with the time topology.
Alternatively, the second method involves directly employing the mass $M$ (%
\ref{a5}) instead. First, we will begin by employing the density (\ref{a13})
and using $M\approx \rho _{\alpha }V_{D-1}$ and (\ref{a6})-(\ref{a13}) in (%
\ref{b3}) we find%
\begin{equation}
\frac{\partial ^{2}\psi \left( \alpha \right) }{\partial \alpha ^{2}}+\frac{%
\alpha \left( E-V\right) }{\alpha _{0}8\pi G_{N}}\sum_{k=2}^{\bar{D}}\left(
\begin{array}{c}
\bar{D} \\
k%
\end{array}%
\right) \left( \pm 1\right) ^{k}\frac{2\pi ^{\frac{\left( D-1\right) }{2}%
}r_{H}^{D-2k-1}\left( \bar{D}\alpha \right) ^{\frac{k-\bar{D}}{\bar{D}-1}}}{%
\left( D-1\right) \Gamma \left( \frac{D-1}{2}\right) }\psi \left( \alpha
\right) =0.  \label{b04}
\end{equation}%
Second, we will substitute the ADM mass (\ref{a5}) in (\ref{b3}). In the
horizon $f(r)=0$, we have%
\begin{eqnarray}
\frac{\partial ^{2}\psi \left( \alpha \right) }{\partial \alpha ^{2}} &=&-%
\frac{2\left( E-V\right) p}{\alpha _{0}\left( D-1\right) }\frac{2\pi ^{\frac{%
\left( D-1\right) }{2}}}{\Gamma \left( \frac{D-1}{2}\right) }r^{D-1}\psi
\left( \alpha \right) \\
&&-2\left( E-V\right) \frac{2\pi ^{\frac{\left( D-1\right) }{2}}\left(
D-2\right) }{16\pi G_{N}\alpha _{0}\Gamma \left( \frac{D-1}{2}\right) }%
\sum_{k=0}^{\bar{D}}\frac{\alpha _{k}\left( \kappa \right) ^{k}r^{D-1}}{%
r^{2k}}\prod\limits_{l=3}^{2k}\left( D-l\right) \psi \left( \alpha \right) .
\end{eqnarray}%
For the case $E=V$, the wave function transforms into a linear equation: $%
\psi \left( \alpha \right) =A\alpha +B$.

\subsection{Stationary solution}

In order to express the solution in line with 4-dimensional Lovelock
gravity, we need to incorporate the relationship between mass and density: $%
M=4\pi r^{3}\rho /3$. From this and (\ref{b3}) we find%
\begin{equation}
\frac{\partial ^{2}\psi \left( \alpha \right) }{\partial \alpha ^{2}}+\frac{%
8\pi r^{3}}{3\alpha _{0}}\left( E-V\right) \rho \psi \left( \alpha \right)
=0.  \label{b4}
\end{equation}%
In the regime of $E=V$, we find $\psi \left( \alpha \right) =A\alpha +B$.
The condition $A=1$ and $B=0$ implies that $\psi \left( \alpha \right)
=\alpha $. If we assume that $r^{3}\left( E-V\right) \rho $ does not depend
on $\alpha $, the normalized ground state wave function in the Einstein
branch can be written as%
\begin{equation}
\psi \left( \alpha \right) =\psi _{1}\cos \left( \sqrt{\frac{16\alpha
^{2}\pi r^{3}}{3\alpha _{0}}\left( E-V\right) \rho }\right) +\psi _{2}\sin
\left( \sqrt{\frac{16\alpha ^{2}\pi r^{3}}{3}\left( E-V\right) \rho }\right)
,  \label{b5}
\end{equation}%
or equivalently%
\begin{equation}
\psi \left( \alpha \right) =\psi _{1}\cos \left( 2\alpha \sqrt{\frac{M}{%
\alpha _{0}}\left( E-V\right) }\right) +\psi _{2}\sin \left( 2\alpha \sqrt{%
\frac{M}{\alpha _{0}}\left( E-V\right) }\right) ,  \label{b6}
\end{equation}%
where $\psi _{1}$ and $\psi _{2}$ are integration constants. This solution
can be expressed as the combination of the function $\psi _{1}\cos \theta $
that travels to the right and the function $\psi _{2}\cos \theta $ that
travels to the left. If $\psi _{1}=\cos \theta _{0}$, $\psi _{2}=\sin \theta
_{0}$ we get $\psi \left( \alpha \right) =\cos \left( 2\alpha \sqrt{\frac{M}{%
\alpha _{0}}\left( E-V\right) }-\theta _{0}\right) $. In our specific
scenario, the quantity $\left\vert \Psi \right\vert ^{2}$ represents the
probability of encountering such a black hole \cite{in2,in3}. For $\psi
_{1}=\psi _{2}$, and using the normalization condition, we get%
\begin{equation}
\psi _{1}=\frac{1}{\sqrt{1+2\cos \left( \sqrt{4\alpha ^{2}\frac{M}{\alpha
_{0}}\left( E-V\right) }\right) \sin \left( \sqrt{4\alpha ^{2}\frac{M}{%
\alpha _{0}}\left( E-V\right) }\right) }}.  \label{b7}
\end{equation}%
Therefore, the wave function takes the following form%
\begin{equation}
\Psi \left( \alpha ,t\right) =\frac{\cos \left( 2\alpha \sqrt{\frac{M}{%
\alpha _{0}}\left( E-V\right) }\right) +\sin \left( 2\alpha \sqrt{\frac{M}{%
\alpha _{0}}\left( E-V\right) }\right) }{\sqrt{1+\sin \left( 4\alpha \sqrt{%
\frac{M}{\alpha _{0}}\left( E-V\right) }\right) }}e^{-iEt}.  \label{b8}
\end{equation}%
For $E=V$\ we get $\Psi \left( \alpha ,t\right) =e^{-iEt}$, this indicates
that the influence of the Lovelock coupling $\alpha $ is negligible in the
limit $E=V$.

\subsection{Solution from topological density}

Thereafter we want to determine the solution of the Schr\"{o}dinger equation
in the context of a 4-dimensional Lovelock branch. We begin by substituting
the 4 dimensional topological density (\ref{a14}) in the given equation (\ref%
{b4}), and we obtain%
\begin{equation}
\frac{\partial ^{2}\psi \left( a\right) }{\partial \alpha ^{2}}+\frac{8\pi
r^{3}}{3\alpha _{0}}\left( E-V\right) \rho _{\alpha }\psi \left( \alpha
\right) =0.  \label{c1}
\end{equation}%
If we are in some regime where $r=r_{H}$, $\kappa =\pm 1$ and $D=4$, Eq. (%
\ref{c1}) can be written as%
\begin{equation}
\frac{\partial ^{2}\psi \left( a\right) }{\partial \alpha ^{2}}+\frac{E-V}{%
6\alpha _{0}G_{N}r_{H}}\alpha \psi \left( \alpha \right) =0.  \label{c2}
\end{equation}%
This equation is similar to the Wheeler-DeWitt equation \cite{AJ1,AJ7} in 4d
quantum gravity. The term $\frac{E-V}{r_{H}}$ represents a force. The
topological solution can be written as%
\begin{equation}
\psi \left( \alpha \right) =\psi _{+}\text{Ai}\left( \left( -\frac{E-V}{%
6\alpha _{0}G_{N}r_{H}}\right) ^{\frac{1}{3}}\alpha \right) +\psi _{-}\text{%
Bi}\left( \left( -\frac{E-V}{6\alpha _{0}G_{N}r_{H}}\right) ^{\frac{1}{3}%
}\alpha \right) ,  \label{c3}
\end{equation}%
where%
\begin{eqnarray}
\text{Ai}\left( \Delta _{H}\alpha \right) &=&\frac{1}{\pi }\int_{0}^{\infty
}\cos \left( \frac{r^{3}}{3}+\alpha r\Delta _{H}\right) dr, \\
\text{Bi}\left( \Delta _{H}\alpha \right) &=&\frac{1}{\pi }\int_{0}^{\infty
}\exp \left( -\frac{r^{3}}{3}+\alpha r\Delta _{H}\right) dr+\frac{1}{\pi }%
\int_{0}^{\infty }\sin \left( \frac{r^{3}}{3}+\alpha r\Delta _{H}\right) dr,
\end{eqnarray}%
where $\Delta _{H}=\left( \frac{V-E}{6\alpha _{0}G_{N}r_{H}}\right) ^{\frac{1%
}{3}}$. The Airy function is derived as the solution to the time-independent
Schr\"{o}dinger equation for a particle confined within a triangular
potential well and for a particle under the influence of a one-dimensional
constant force field. Additionally, it is useful for providing consistent
semiclassical approximations near a turning point within the WKB
approximation, where the potential can be locally approximated by a linear
function of position. The solution related to the triangular potential well
is particularly relevant for understanding electrons confined within
semiconductor heterojunctions. Thus, the wave function is expressed as
follows.
\begin{equation}
\Psi \left( \alpha ,t\right) =\psi _{+}e^{-iEt}\text{Ai}\left( \left( -\frac{%
E-V}{6G_{N}\alpha _{0}r_{H}}\right) ^{\frac{1}{3}}\alpha \right) +\psi
_{-}e^{-iEt}\text{Bi}\left( \left( -\frac{E-V}{6\alpha _{0}G_{N}r_{H}}%
\right) ^{\frac{1}{3}}\alpha \right) .  \label{c6}
\end{equation}%
The normalization condition states that the probability of the position x
falling within the interval $\alpha _{1}\leq \alpha \leq \alpha _{2}$ is
given by the integral of the density over this interval: $P_{\alpha
}=\int_{\alpha _{1}}^{\alpha _{2}}\left\vert \Psi \left( \alpha ,t\right)
\right\vert ^{2}d\alpha $. \cite{AJ6}. By denoting $t$ as the time when the
particle was detected, we arrive at the normalization condition: $%
\int_{-\infty }^{+\infty }\left\vert \Psi \left( \alpha ,t\right)
\right\vert ^{2}d\alpha =1$ where $\left\vert \Psi \left( \alpha ,t\right)
\right\vert ^{2}=\int_{-\infty }^{\infty }d\alpha \Psi \Psi ^{\ast }$. Note
that for $\Psi _{0}>0$, the wave function will oscillate around zero, and
the amplitude will be controlled by the value of a. For $\Psi _{0}<0$, the
wave function will be inverted (mirror image) with respect to the $\alpha $%
-axis.
\begin{figure}[H]
\centering\includegraphics[width=11cm]{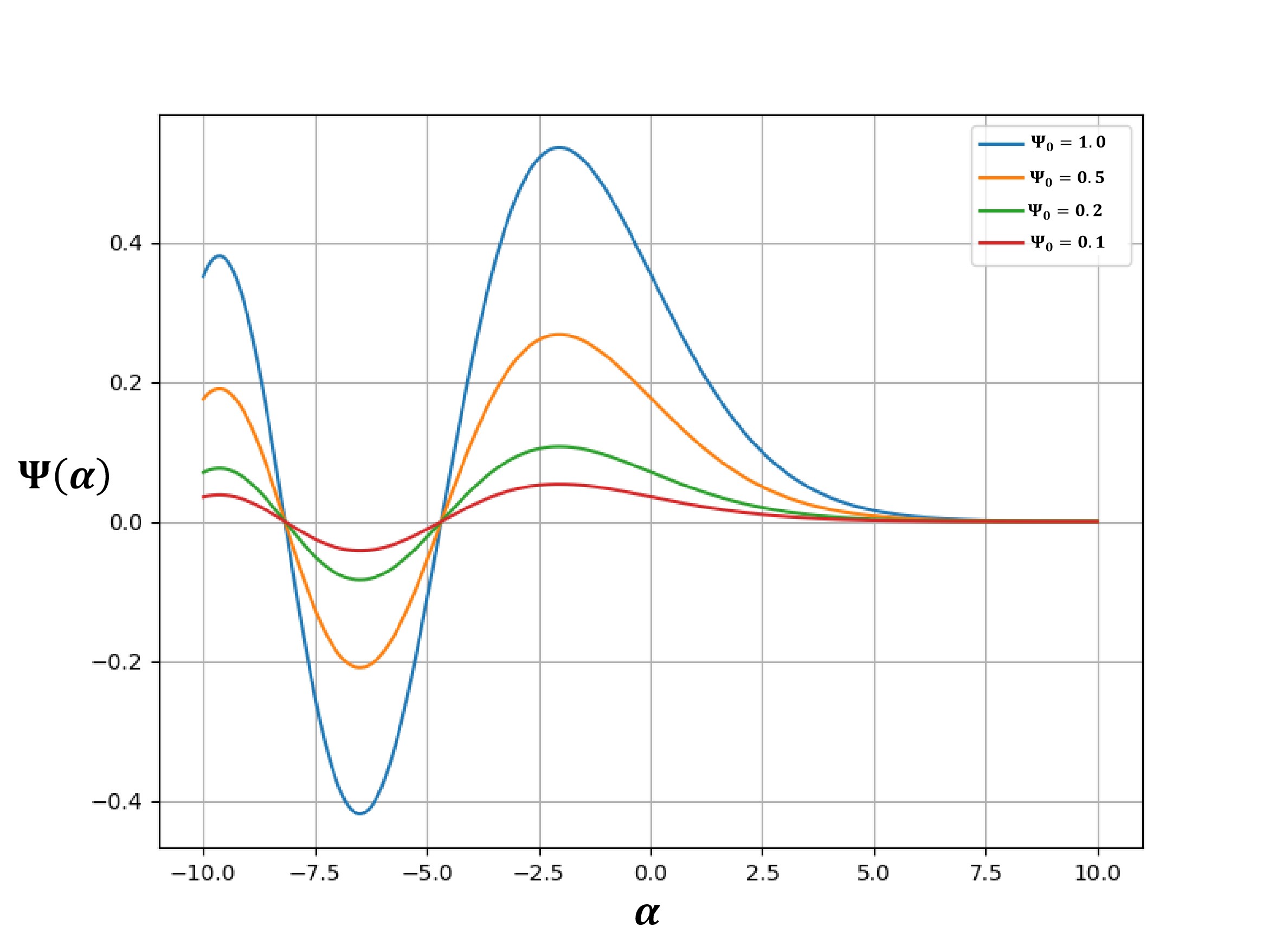}
\caption{The figure displays the wave function $\Psi _{+}\left( \protect%
\alpha \right) $ for different values of the scaling factor $\Psi _{0}$ and
a constant value of $\Delta _{H}=0.5$. The wave function is defined using
the Airy function of the first kind. We have $\Psi _{0}=\protect\psi %
_{+}e^{-iEt_{0}}$ and $\Delta _{H}=\left( \frac{V-E}{6\protect\alpha %
_{0}G_{N}r_{H}}\right) ^{\frac{1}{3}}$.}
\label{F2}
\end{figure}
\begin{figure}[H]
\centering\includegraphics[width=11cm]{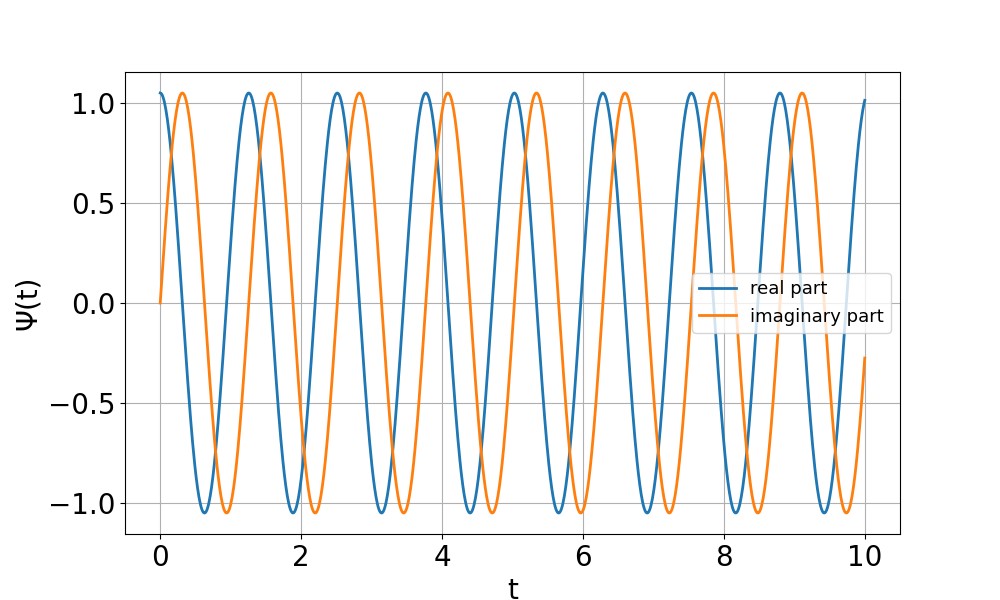}
\caption{The figure displays the wave function $\Psi \left( t\right) $ for
different values of $\protect\psi _{+}=\protect\psi _{-}=1$, $\protect\alpha %
=0.1,$ $E=-5,$ $6G_{N}\protect\alpha _{0}r_{H}=1$ and $V=0.4$.}
\label{F3}
\end{figure}
\begin{figure}[H]
\centering\includegraphics[width=11cm]{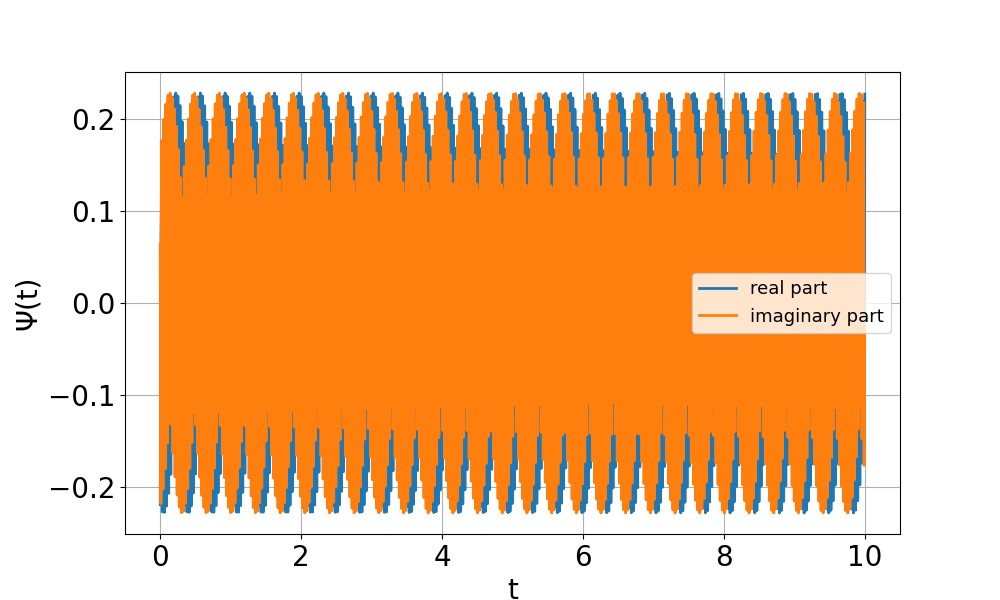}
\caption{The figure displays the wave function $\Psi \left( t\right) $ for
different values of $\protect\psi _{+}=\protect\psi _{-}=1$, $\protect\alpha %
=0.1,$ $E=5\times 10^{4},$ $6G_{N}\protect\alpha _{0}r_{H}=1$ and $V=0.4$.}
\label{F4}
\end{figure}
\begin{figure}[H]
\centering\includegraphics[width=11cm]{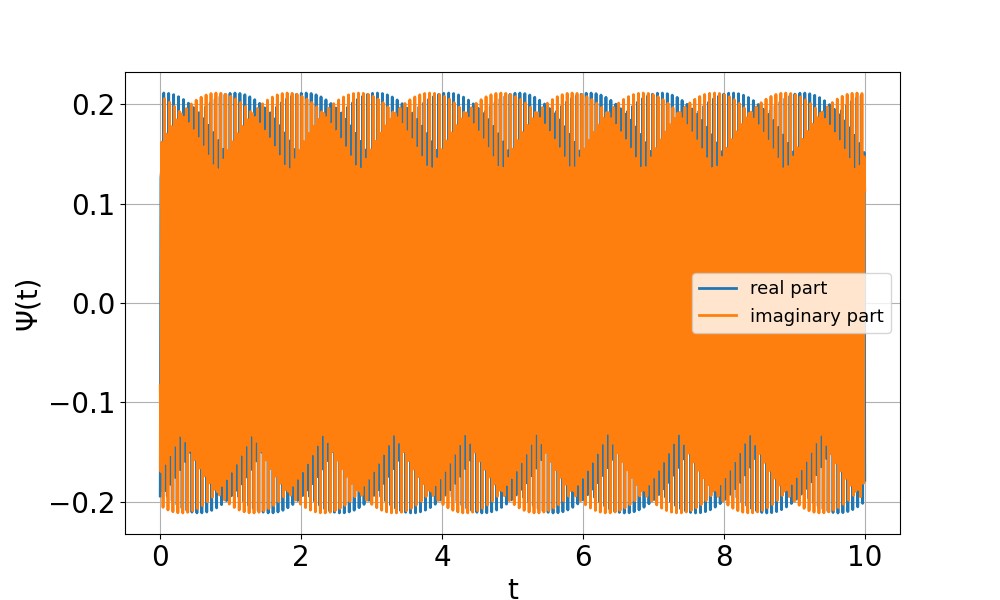}
\caption{The figure displays the wave function $\Psi \left( t\right) $ for
different values of $\protect\psi _{+}=\protect\psi _{-}=1$, $\protect\alpha %
=0.1,$ $E=5\times 10^{6},$ $6G_{N}\protect\alpha _{0}r_{H}=1$ and $V=0.4$.}
\label{F5}
\end{figure}

\section{Conclusion}

This paper explores solutions to the Klein-Gordon equation and Schr\"{o}%
dinger equation with a wave function within the framework of Lovelock
gravity. Our findings establish a significant connection between the
Lovelock coupling and the Hubble parameter, infusing the coupling with
profound physical implications. Additionally, we have presented the Smarr
formula derived from the topological density. These outcomes offer valuable
insights into the interaction between quantum wave functions and
gravitational theories, elucidating the intricate relationship between
Lovelock gravity and the behavior of the universe on cosmological scales.
Notably, one of the Klein-Gordon solutions indicates the presence of the
wave function near the horizon. After introducing the Wheeler-de Witt
Hamiltonian, we proceeded to compute the solutions for scalar fields when $%
D\rightarrow 4$. Furthermore, employing spherical harmonics, we examined the
relationship between a black hole's temperature and its potential. By
analyzing static and spherically symmetric geometry, we have identified a
significant link between the potential of the spherical harmonics and the
the tortoise coordinate. This association not only provides insights into
the behavior of gravitational fields but also defines the Schwarzschild
limit. By utilizing Schr\"{o}dinger's equation, we have shown that the Airy
function Ai presents a viable solution. We have established a significant
relationship between the Lovelock coupling and the horizon radius. This
result suggests that the horizon possesses two discretions: geometric and
topological.

\section{Appendix A}

Firstly using the Schr\"{o}dinger equation of a wave function $\Psi =\Psi
\left( \alpha ,t\right) $\ around the horizon ($r\sim r_{H}$)%
\begin{equation}
i\frac{\partial \Psi }{\partial t}=-\frac{1}{2M}\frac{\partial ^{2}\Psi }{%
\partial r^{2}}+V\Psi .
\end{equation}%
Regarding the equivalence of $\alpha $\ and $r$\ in the horizon in (\ref{ac2}%
): $r=\frac{\alpha }{\sqrt{\alpha _{0}}}$, i.e.%
\begin{equation*}
\frac{\partial ^{2}}{\partial r^{2}}=\alpha _{0}\frac{\partial ^{2}}{%
\partial \alpha ^{2}},
\end{equation*}%
\ Consequently, we get Eq. (\ref{b1})%
\begin{equation}
i\frac{\partial \Psi }{\partial t}=-\frac{\alpha _{0}}{2M}\frac{\partial
^{2}\Psi }{\partial \alpha ^{2}}+V\Psi .
\end{equation}%
We consider a stationary state with energy $E$\ as
\begin{equation*}
\Psi \left( \alpha ,t\right) =\psi \left( \alpha \right) e^{-iEt}
\end{equation*}%
we obtain%
\begin{equation}
i\psi \left( \alpha \right) \frac{\partial e^{-iEt}}{\partial t}=-\frac{%
\alpha _{0}}{2M}e^{-iEt}\frac{\partial ^{2}\psi \left( \alpha \right) }{%
\partial \alpha ^{2}}+V\psi \left( \alpha \right) e^{-iEt}.
\end{equation}%
So, the Schr\"{o}dinger stationary equation is
\begin{equation}
\frac{\partial ^{2}\psi \left( \alpha \right) }{\partial \alpha ^{2}}+2\frac{%
M}{\alpha _{0}}\left( E-V\right) \psi \left( \alpha \right) =0.
\end{equation}

\end{document}